\documentstyle[12pt]{article}
\begin{document}
\begin{titlepage}

\vspace{2.5cm}
\begin{centering}
{\LARGE{\bf Casimir effect around\\}}
\vspace{0.5cm}
{\LARGE {\bf a screw dislocation}}\\
\bigskip\bigskip
Ivan Pontual and Fernando Moraes\\
{\em Departamento de F\'{\i}sica\\
Universidade Federal de Pernambuco\\
50670-901 Recife, PE, Brazil}\\
\vspace{1cm}
\end{centering}
\vspace{1.5cm}

\begin{abstract}
In this work, it is shown that a non-zero vacuum energy density (the Casimir energy) for a scalar field appears in a continuous elastic solid due to the presence of a topological defect, the screw dislocation. An exact expression is obtained for this energy density in terms of the Burgers vector describing the defect, for zero and finite temperature. 
\end{abstract}

\end{titlepage}

\section{Introduction}

The importance of the Casimir effect~\cite{Cas,Plu,Mos} in nowadays physics can hardly
be overemphasized. It lies at the core of the modern definition of vacuum
energy of quantum fields. Ideally, these fields are placed in an unbounded,
infinite Minkowski spacetime. In practice, however, they may undergo
interactions, which can be viewed as suitable boundary conditions on them,
or they may be defined in a background spacetime of non-trivial topology
(presence of defects and boundaries) or geometry (curvature and/or
torsion). This causes variations in the zero-point energy of the quantum
fields, and the difference between the modified energy and the Minkowskian
one is the so-called Casimir energy (see, e.g., Ref. ~\cite{Plu} for examples.).

The Casimir effect in the presence of conical topologies has been
extensively studied in the context of cosmic strings~\cite{Vil,Mor,Lin,Dow}, but the
applications in condensed matter systems have remained quite unexplored.
Topological defects in solids (disclinations and dislocations) offer a
natural setting for vacuum polarization to occur, as pointed out in a recent
paper by one of us (F.M.)~\cite{Fer}. It was studied therein the Casimir effect in
the presence of a disclination. The natural step further is to extend such
results to the dislocation case, as these are more common in a solid~\cite{Kit}.

In this work we calculate the Casimir energy for a scalar field in the
presence of a screw dislocation, which is done by studying this field in a
spacetime endowed with a non-Euclidean metric simulating the defect~\cite{Tod}. The reason as to why we deal with this defect is that it proved
to be the most tractable one after the disclination case, and also because
it is related to the still unsolved problem of the Casimir effect in the
presence of a rotating cosmic string ~\cite{Mor,Mat}, which brings naturally
cosmology and solid state physics together formally, for mutual benefit.
The scalar field is the simplest field and can be immediately generalized to
the electromagnetic and phonon fields cases by just decomposing these along
their polarization axis. Also, in Ref. ~\cite{Fer} it was argued that fermionic
fields (electrons in this case) should have a much smaller associated
Casimir effect than the electromagnetic field due to an exponential damping
related to the electronic mass.

This paper is organized as follows: in Sec.2 we review briefly the $\zeta $%
-function regularization technique as introduced by Hawking in Ref. ~\cite{Haw}. We
then use it in Sec. 3 to obtain an expression for the contribution of the
vacuum energy due to the defect. In Sec. 4 we derive an asymptotic expression for large distances, useful for practical purposes. In Sec. 5 we present our concluding remarks. In Appendix A we give a brief definition of the heat kernel of an operator and its connection with the generalized $\zeta $-function, for the benefit of the reader unfamiliar with these concepts. The heat kernel for the screw dislocation is obtained in Appendix B. Throughout the paper we use units in which $c=\hbar =k_B=1$.

\section{$\zeta $-function regularization}

The $\zeta $-function technique ~\cite{Haw,Cri} is a powerful tool to calculate the
Casimir energy. Among its advantages we have that the effects due to
temperature may be implemented in a very straightforward manner, via the
Euclidean time formalism ~\cite{Kle}. As we have a spatial defect, we shall be
concerned with an {\it ultrastatic }spacetime manifold ${\cal M}$, i.e., $%
{\cal M}$ $=R\times \Sigma $, where $\Sigma $ is $R^{3}$ modified by the
defect. We may perform a Wick rotation of the time coordinate $x^{0}$%
 to the imaginary time $\tau =ix^0$, and impose that the scalar field $%
\varphi $ be periodic in imaginary time with period $\beta =\frac 1T$ , the
inverse temperature. In this context, the partition function for the scalar
field with mass $m$ at temperature $T$ is defined as (our notation and
outlook follow ~\cite{Van}): 
\begin{equation}
Z_\beta =\int_{\varphi \left( 0,x\right) =\varphi \left( \beta ,x\right) }%
{\cal D}\varphi \exp \left( -\frac 12\int_{{\cal M}}\varphi L_4\varphi
d^4x\right) =\exp [\frac 12\zeta _\beta ^{\prime }(0\mid \frac{L_4}{\mu ^2}%
)],
\end{equation}
where $\mu $ is an arbitrary renormalization parameter from the path
integral measure, $L_4=-\Delta _{LB}^{\left( 4\right) }+m^2$, and we
represented by $\zeta _\beta ^{\prime }(0\mid \frac L{\mu ^2})$ the
derivative at zero of the (generalized) $\zeta $-function of the operator $%
L_4$ (see Appendix A). After a Wick rotation the topology of ${\cal M}$
becomes $S^1\times \Sigma $, and $\Delta _{LB}^{\left( 4\right) }$%
stands for the four dimensional Laplace-Beltrami operator for the
(Euclidianized) metric on ${\cal M}$: 
\begin{equation}
\Delta _{LB}^{\left( 4\right) }=\frac 1{\sqrt{g}}\partial _\mu \left[ \sqrt{g%
}g^{\mu \nu }\partial _\nu \right] ,
\end{equation}
where $g_{\mu \nu }$ is the metric tensor on ${\cal M}=S^1\times \Sigma $,
and $g$ is its determinant. For Wick-rotated ultrastatic manifolds, one has
that the LB operator can be decomposed as $\Delta _{LB}^{\left( 4\right)
}=-\partial _\tau -\Delta _{LB}^{\left( 3\right) }$, and now $\Delta
_{LB}^{\left( 3\right) }$ is the LB operator for the metric on $\Sigma $
(spatial part of the metric on ${\cal M}$).

Eq. (1) provides all relevant regularized physical quantities via the
usual formulas of thermodynamics, with one difference from the ordinary
partition function: the presence of a vacuum term, which gives the
zero-temperature contribution.This can be isolated in the limit $\beta
\rightarrow \infty $. By denoting $L_3=-\Delta _{LB}^{(3)}+m^2$, one can use
the relation between the heat kernel of an operator and its $\zeta $%
-function (Appendix A) and some algebraic manipulations to obtain ~\cite{Cog} : 
\begin{eqnarray}
\zeta (s\mid \frac{L_4}{\mu ^2})&=&\frac{\mu \beta \Gamma \left( s-1/2\right) 
}{\sqrt{4\pi }\Gamma \left( s\right) }\zeta (s-1/2\mid \frac{L_3}{\mu ^2})+%
\frac{\mu \beta }{\sqrt{\pi }\Gamma \left( s\right) }\sum_{n=1}^\infty
\int_0^\infty t^{s-1/2}\times\nonumber\\[0.3cm]
&&e^{-(n\mu \beta )^2/4t}Tr\ e^{-tL_3/\mu ^2}dt,
\end{eqnarray}
from which it is easily deduced [15,16], that 
\begin{equation}
\ln Z_\beta =-\frac \beta 2\zeta (-1/2\mid L_3)+\frac \beta {\sqrt{4\pi }%
}\sum_{n=1}^\infty \int_0^\infty t^{-3/2}e^{-(n\beta )^2/4t}Tr\ e^{-tL_3}dt,
\end{equation}
in the case one has scale independence of the Casimir energy \footnote{%
The Casimir energy may depend on the scale $\mu $ and in this case the expression is
more complicated. In Ref.~\cite{Van} it is displayed the general formula. In our
case, and in the majority of the physically relevant systems, one has scale
independence, and therefore we omit further comments on this. For a discussion of these issues, see Ref.~\cite{Bla}.}. In Eqs. (3)
and (4), $Tr\ e^{-tL_3}$ stands for the trace of the heat kernel of $L_3$
(Appendix A). We shall be interested in the free energy of the field $F=-\ln
Z_\beta /\beta $, which immediately yields 
\begin{equation}
F=\frac 12\zeta (-1/2\mid L_3)-\frac 1{\sqrt{4\pi }}\sum_{n=1}^\infty
\int_0^\infty t^{-3/2}e^{-(n\beta )^2/4t}Tr\ e^{-tL_3}dt.
\end{equation}
This last equation shows clearly the splitting between the $T=0$ term and
the contribution from thermal excitations, the latter being identified with
the ordinary thermodynamical free energy. One has hereby a conceptually
simple method to calculate the Casimir energy, just by obtaining the heat
kernel of $L_3$ and its $\zeta $-function. We proceed to do this in next
section, for the explicit case of a screw dislocation.

\section{Casimir energy around a screw dislocation}

In this section we basically use a variant of Eq. (5) to obtain an
expression for the Casimir energy. We follow the theory of
defects/three-dimensional gravity of Katanaev-Volovich ~\cite{Kat} to model a screw
dislocation by endowing the elastic infinite continuum representing the
solid with a metric given, in cylindrical coordinates, by ~\cite{Tod}: 
\begin{equation}
ds^2=dr^2+r^2d\phi ^2+\left( dz+\kappa d\phi \right) ^2,
\end{equation}
where $0\leq \phi \leq 2\pi $. This metric represents an infinite defect
along the z-axis, resulting from ``cutting '' the space along the semiplane $%
\phi =0$, $r\geq 0$, and ``sliding '' it along the z axis by an amount $b$,
the modulus of the Burgers vector. This process gives a helix-like topology to the
space, the step of which is $\kappa \equiv \frac b{2\pi }$. In the case $b=0$
we obviously recover the flat Euclidean metric, describing the medium in the
absence of dislocations. The metric in Eq. (6) is locally flat, as can be easily seen by performing the coordinate transformation $Z^{\prime
}\equiv z+\kappa \phi $: 
\begin{equation}
ds^2=dr^2+r^2d\phi ^2+dZ^{^{\prime }2},
\end{equation}
which is flat everywhere except at $r=0$, where it has a $\delta $-function
singularity in the torsion ~\cite{Tod}.One may infer the helix-like behavior by
noting that the periodicity of $\phi $ induces the identification: 
\begin{equation}
(t,r,\phi ,Z^{\prime })\sim (t,r,\phi +2\pi ,Z^{\prime }+2\pi \kappa ).
\end{equation}

As we mentioned before, our interest is in deriving the heat kernel of the
operator $L_3$ associated with this metric. This is done in Appendix B, and
the result is: 
\begin{eqnarray}
K(\vec{r},\vec{r}^{\,\prime },t)&=&\frac{e^{-m^2t}e^{-\frac{%
(r^2+r^{^{\prime }2})}{4t}}}{2(2\pi )^2t}\sum_{n=-\infty }^\infty
e^{-in(\phi -\phi ^{\prime })}\int_{-\infty }^\infty e^{-\nu ^2t}e^{i\nu
(z-z^{\prime })}\times\nonumber\\[0.3cm]
&&I_{\mid n+\kappa \nu \mid }\left( \frac{rr^{\prime }}{2t}%
\right) d\nu , 
\end{eqnarray}
where $I_p(z)$ is the modified Bessel function of the first kind. If we
consider $\vec{r}=\vec{r}^{\,\prime }$, Eq.(9) becomes: 
\begin{equation}
K(\vec{r},\vec{r},t)=\frac{e^{-m^2t}e^{-\frac{r^2}{2t}}}{2(2\pi
)^2t}\sum_{n=-\infty }^\infty \int_{-\infty }^\infty e^{-\nu ^2t}I_{\mid
n+\kappa \nu \mid }\left( \frac{r^2}{2t}\right) . 
\end{equation}

The so-called local $\zeta $-function, which plugged into Eq. (5) gives the
free energy density (see Appendix A), is obtained by setting $\vec{r}=%
\vec{r}^{\,\prime }$ in Eq. (9) and applying a Mellin transform
~\cite{Gra}: 
\begin{equation}
\zeta (s,\vec{r})\equiv \frac 1{\Gamma \left( s\right) }\int_0^\infty
t^{s-1}K(\vec{r},\vec{r},t)dt. 
\end{equation}

Before performing the Mellin transformation, however, we shall work Eq. (10)
out to reshape it into a more convenient form. Let us consider the well
known integral representation ~\cite{Gra}: 
\begin{equation}
I_\nu \left( z\right) =\frac 1{2\pi }\int_{-\pi }^\pi e^{z\cos \theta }\cos
\nu \theta \ d\theta -\frac{\sin \nu \pi }\pi \int_0^\infty e^{-z\cosh x-\nu
x}dx,
\end{equation}
valid for $Re\ z>0$ and $Re\ \nu \geq 0$. We substitute it into Eq. (10) to obtain 
\begin{eqnarray}
K(\vec{r},\vec{r},t)&=&\frac{e^{-m^2t}e^{-\frac{r^2}{2t}}}{2(2\pi
)^2t}\sum_{n=-\infty }^\infty [\frac 1{2\pi }\int_{-\pi }^\pi d\theta \ e^{%
\frac{r^2}{2t}\cos \theta }\int_{-\infty }^\infty d\nu \ e^{-\nu ^2t}\times\nonumber\\[0.3cm]
&&\cos\left[ \left( n+\kappa \nu \right) \theta \right] -\frac 1\pi \int_{-\infty }^{\infty} d\nu \ e^{-\nu ^2t}\sin {}{}\!\mid
n+\kappa \nu \mid \times \nonumber\\[0.3cm]
&&\int_0^\infty dx\ e^{-\frac{r^2}{2t}\cosh x}e^{-\mid
n+\kappa \nu \mid x}].
\end{eqnarray}
We shall deal separately with the integrals. We have that ~\cite{Gra}: 
\begin{equation}
\int_{-\infty }^\infty d\nu \ e^{-\nu ^2t}\cos \left[ \left( n+\kappa \nu
\right) \theta \right] =\sqrt{\frac \pi t}e^{-\frac{\kappa ^2\theta ^2}{4t}%
}\cos (n\theta ).
\end{equation}

We use here the Poisson summation formula ~\cite{Plu}: 
\begin{equation}
\sum_{n=-\infty }^\infty F(n)=2\pi \sum_{n=-\infty }^\infty c\left( 2\pi
n\right) ,
\end{equation}
where 
\begin{equation}
c\left( \alpha \right) =\frac 1{2\pi }\int_{-\infty }^\infty dx\ e^{i\alpha
x}F\left( x\right) .
\end{equation}
This easily gives the well-known formula: 
\begin{equation}
\sum_{n=-\infty }^\infty \delta \left( x-n\right) =\sum_{n=-\infty }^\infty
e^{i2\pi nx},
\end{equation}
which provides the following representation: 
\begin{equation}
\sum_{n=-\infty }^\infty \cos \left( n\theta \right) =\frac 12\left[
\sum_{n=-\infty }^\infty \delta \left( \frac \theta {2\pi }-n\right)
+\sum_{n=-\infty }^\infty \delta \left( \frac \theta {2\pi }+n\right)
\right] .
\end{equation}
We recall that $\delta \left( \lambda x\right) =\lambda ^{-1}\delta \left(
x\right) $, and upon substitution of Eq.(14) into Eq.(13), and then substituting Eq. (18) into Eq. (13), we obtain: 
\begin{eqnarray}
K(\vec{r},\vec{r},t)&=&\frac{e^{-m^2t}}{(4\pi t)^{3/2}}-\frac{%
e^{-m^2t}e^{-\frac{r^2}{2t}}}{8\pi ^3t}\sum_{n=-\infty }^\infty
\int_{-\infty }^\infty d\nu \ e^{-\nu ^2t}\times \nonumber \\[0.3cm] 
&&\sin \pi \!\mid n+\kappa \nu \mid \int_0^\infty dx\ e^{-\frac{r^2}{2t}%
\cosh x}e^{-\mid n+\kappa \nu \mid x}.
\end{eqnarray}
The next step is to calculate the local $\zeta $-function (see Appendix A): 
\begin{equation}
\zeta (\vec{r},s)=\frac 1{\Gamma \left( s\right) }\int_0^\infty
t^{s-1}K(\vec{r},\vec{r},t)dt,
\end{equation}
which by means of a simple representation of the modified Bessel function of
the second kind, 
\begin{equation}
2\left( \frac \alpha {\alpha ^{\prime }}\right) ^{\nu /2}K_\nu \left( 2\sqrt{%
\alpha \alpha ^{\prime }}\right) =\int_0^\infty dt\ t^{\nu -1}\exp \left(
-\frac \alpha t-\alpha ^{\prime }t\right) ,
\end{equation}
may be written as: 
\begin{eqnarray}
\zeta (\vec{r},s)&=&\frac{\Gamma \left( s-3/2\right) m^{-2s+3}}{(4\pi
)^{3/2}\Gamma \left( s\right) }-\frac 1{4\pi ^3\Gamma \left( s\right)
}\int_{-\infty }^\infty d\nu \ \int_0^\infty dx\ F_b\left( \nu ,x\right)
\times\nonumber  \\ [0.3cm]
&& \left( \frac{r\cosh \left( \frac x2\right) }{\sqrt{\nu ^2+m^2}}%
\right) ^{s-1}K_{s-1}(2r\cosh \left( \frac x2\right) \sqrt{\nu ^2+m^2}),
\end{eqnarray}
where we have defined 
\begin{equation}
F_b\left( \nu ,x\right) =\sum_{n=-\infty }^\infty \sin \pi \!\mid n+\kappa
\nu \mid e^{-\mid n+\kappa \nu \mid x}.
\end{equation}
We may notice that for $\kappa =0$ this last summation vanishes. Therefore,
in the absence of the defect ($b=0$), we see that only the first term on the
right hand side of Eq.(22) remains. By substituting the Euclidean heat
kernel, Eq.(B.7) of Appendix B, into Eq.(11), we see that this term corresponds to the free
space $\zeta $-function. The Casimir energy is renormalized with respect to
the flat, infinite Euclidean space, and we shall accordingly disregard the
Euclidean term and thus obtain the defect $\zeta $-function: 
\begin{eqnarray}
\zeta _b(\vec{r},s)&=&-\frac 1{4\pi ^3\Gamma \left( s\right)
}\int_{-\infty }^\infty d\nu \ \int_0^\infty dx\ F_b\left( \nu ,x\right)
\times \nonumber \\[0.3cm] 
&&\times \left( \frac{r\cosh \left( \frac x2\right) }{\sqrt{\nu ^2+m^2}}%
\right) ^{s-1}K_{s-1}(2r\cosh \left( \frac x2\right) \sqrt{\nu ^2+m^2}).
\end{eqnarray}

As it stands, we have the Casimir energy density at zero temperature: it is
simply $E=\frac 12\zeta _b(\vec{r},s)$. We may get the full energy,
valid for any temperature by considering Eq.(A.11) of the Appendix A. It is
clear that our problem is to calculate the integral : 
\begin{equation}
I_p=\int_0^\infty t^{-3/2}K(\vec{r},\vec{r},t)e^{-\frac{\left(
p\beta \right) ^2}{4t}}dt, 
\end{equation}
for each natural number $p$. The procedure is entirely analogous to the
former calculation, and it yields: 
\begin{eqnarray}
I_p&=&-\frac 1{4\pi ^3}\int_{-\infty }^\infty d\nu \ \int_0^\infty dx\
F_b\left( \nu ,x\right)\left[ \sqrt{\frac{r^2\cosh ^2\left( \frac x2\right) +\frac{\left(
p\beta \right) ^2}4}{\nu ^2+m^2}}\,\right] ^{-3/2}\times\nonumber\\[0.3cm]
&&K_{-3/2}(2\sqrt{r^2\cosh
^2\left( \frac x2\right) +\frac{\left( p\beta \right) ^2}4}\sqrt{\nu ^2+m^2}%
).
\end{eqnarray}

Then, we finally have the free energy density for a massive scalar field in
the presence of a screw dislocation at any temperature: 
\begin{eqnarray}
{\cal F}&=&\frac 12\zeta _b(\vec{r},-1/2)+\frac 1{8\pi
^3}\sum_{p=1}^\infty \int_{-\infty }^\infty d\nu \ \int_0^\infty dx\
F_b\left( \nu ,x\right) \left[ \sqrt{\frac{r^2\cosh ^2\left( \frac x2\right) +\frac{\left(
p\beta \right) ^2}4}{\nu ^2+m^2}}\,\right] ^{-3/2}\times\nonumber\\[0.3cm]
&&K_{-3/2}(2\sqrt{r^2\cosh
^2\left( \frac x2\right) +\frac{\left( p\beta \right) ^2}4}\sqrt{\nu ^2+m^2}%
).
\end{eqnarray}
The massless case may also be otained by directly setting $m=0$ in this
equation. The value $\zeta _b(\vec{r},-1/2)$ should be understood in
the following way: one assumes that $s$ is large enough to make the
integrals in Eq.(22) converge, performs the integration in terms of $s$ and
afterwards one analitically continues the resulting expression to $s=-1/2$.
This cannot be done explicitly in our case due to the complexity of the
expressions. Thus we have derived an exact expression for the Casimir
energy, but because it is a complicated one, in the next section we shall
derive an asymtotic formula for the simplest particular case, namely the
massless field at zero temperature.

\section{Asymptotic expression}

Our problem here amounts to see what our formulas become if we take $\frac{%
rr^{\prime }}{2t}\rightarrow \infty $. From the physical standpoint, one
expects that the qualitative features of the energy functional are preserved
at large distances, since the defect is axially symmetric and infinite. On
the other hand, our considering the solid as a continuum medium is an
approximation valid for large distances from the defect.

Let us now consider Eq.(10) for the heat kernel. We set $m=0$ and and take
the following asymptotic expression for the modified Bessel function for $%
\mid w\mid \rightarrow \infty $ ~\cite{Gra}: 
\begin{equation}
I_{\mid \lambda \mid }\left( w\right) \approx \left( 2\pi w\right)
^{-1/2}\exp \left( w-\frac 1{2w}\left( \lambda ^2-1/4\right) \right) .
\end{equation}

Now, we just set $w=\frac{r^2}{2t}$ and $\lambda =n+\kappa \nu $, and
substitute Eq.(28) into Eq.(10).The result is: 
\begin{equation}
K(\vec{r},\vec{r},t)=\frac{e^{\frac t{4r^2}}}{8\pi ^2\sqrt{\pi t}r}%
\sum_{n=-\infty }^\infty e^{-\frac{n^2t}{r^2}}\int_{-\infty }^\infty d\nu \
e^{-t(1+\frac{\kappa ^2}{r^2})\left( \nu ^2+\frac{2\kappa \nu n}{r^2\left( 1+%
\frac{\kappa ^2}{r^2}\right) }\right) }.
\end{equation}
We complete the square in the last exponencial and, defining a new variable $%
u\equiv \nu +\frac{\kappa n}{r^2\left( 1+\frac{\kappa ^2}{r^2}\right) }$,
integrate with respect to $u$ to obtain: 
\begin{equation}
K(\vec{r},\vec{r},t)=\frac{e^{\frac t{4r^2}}}{8\pi ^2tr}\ (1+\frac{%
\kappa ^2}{r^2})^{-1/2}\sum_{n=-\infty }^\infty e^{-\frac t{r^2}n^2}\ e^{t%
\frac{n^2\kappa ^2}{r^2\left( 1+\frac{\kappa ^2}{r^2}\right) }}.
\end{equation}
Thus, we simply expand $e^{\frac t{4r^2}}\approx 1+\frac t{4r^2}$ and apply
the Mellin transform, which yields the local zeta function for the massless
field: 
\begin{eqnarray}
\zeta _{\left( m=0\right) }\left( s,\vec{r}\right) &=&\frac{\Gamma \left(
s-1\right) }{4\pi ^2\Gamma \left( s\right) }r^{2s-3}\ (1+\frac{\kappa ^2}{r^2%
})^{s-3/2}\zeta _R\left( 2s-2\right) +\nonumber\\[0.3cm]
&&\frac 1{16\pi ^2\Gamma \left( s\right)
}r^{2s-3}\ (1+\frac{\kappa ^2}{r^2})^{s-3/2}\zeta _R\left( 2s\right) ,
\end{eqnarray}
where we have substituted $\sum_{n=1}^\infty n^{-z}\equiv \zeta _R\left(
z\right) $, the Riemann zeta function ~\cite{Gra}. Now we put $s=-1/2$ and pick the
values of the Riemann zeta function and gamma function in any standard table
of special functions, e.g., Ref. ~\cite{Gra}, then we finally have the simple
expression: 
\begin{equation}
\zeta _{\left( m=0\right) }\left( -1/2,\vec{r}\right) =-\frac 1{720\pi
^2}\ (1+\frac{\kappa ^2}{r^2})^{-2}\frac 1{r^4}-\frac 1{480\pi ^2}\ (1+\frac{%
\kappa ^2}{r^2})^{-1}\frac 1{r^4}.
\end{equation}

We have seen that we must subtract the Euclidean contribution in order to
obtain the renormalized Casimir energy density: 
\begin{equation}
{\cal E}_{ren.}=\frac 12[\zeta _{\left( m=0\right) }\left( -1/2,\vec{r}%
\right) -\zeta _{\left( m=0\right) }\left( -1/2,\vec{r}\right) \mid
_{\kappa =0}]. 
\end{equation}

From this expression we obtain, for instance, the lowest order term of an
expansion in powers of $\frac{\kappa ^2}{r^2}$: 
\begin{equation}
{\cal E}_{ren.}=\frac 7{2880\pi ^2}\frac{\kappa ^2}{r^6}+{\cal O}\left( 
\frac{\kappa ^4}{r^8}\right) . 
\end{equation}
Note that in the units we use here the energy has units of (lenght)$^{-1}$
That the sign is positive should already be expected, since at $T=0$ the
addition of the defect to the background increases the energy of the field,i.e., the favoured
configuration is $b=0$, the flat space without defects. Only even powers of $%
\kappa $ occur, since it is immaterial from an energetic viewpoint whether
the ``helix'' is right-handed ($b$ positive) or left-handed ($b$ negative).

\section{Concluding Remarks}

In the present work we give an example of the importance of quantum field theoretical effects in solids with defects. Albeit very simplified, our model for the solid (infinite, non-magnetic and with only one defect), gives already a non-zero Casimir energy, which can play a role in various phenomena in solid state physics, for instance, contributing to the specific heat of solids.

It would be interesting to investigate an extension of this model to magnetic materials. In the continuous approximation of the Ising model, for instance, this amounts to add a term proportional to $\varphi^{4}$~\cite{Ami} to the argument of the first exponential in Eq.(1). This model would no longer yield an exact result but could be handled perturbatively.

Another line of interest is to consider classical effects of the defect. Specifically, it is known~\cite{Cla} that a classical point charge in the presence of a disclination suffers a self force induced by the defect. An identical effect would be expected in the dislocation case, and possibly give rise to bound states. We are presently tackling this problem.
\\
\\
{\large{\bf Appendix A: $\zeta $-function and the heat kernel operator}}
\\
\\
The $\zeta $-function technique as used in this paper was sistematically
applied for the first time in Ref. ~\cite{Cri}, but Hawking ~\cite{Haw} shaped it into its
present form. We briefly introduce the subject in this Appendix. For a more
detailed account we suggest Ref. ~\cite{Haw} and the comprehensive (and very
complete) book by Elizalde {\it et al}. ~\cite{Eli}.

One starts with the description of a quantum physical system by a
Hamiltonian $\hat{H}$, with adequate boundary conditions and, possibly,
non-trival background metric and topology. Mathematically, this reduces to
an elliptic, self-adjoint, second order differential operator, with
correspondent boundary conditions. Generally, it is not possible to
calculate its spectrum directly, but we can still define, for such a kind of
operator, the (generalized) zeta function of $\hat{H}$, given by:
$$\zeta _{\hat{H}}\left( s\right) \equiv Tr\exp \left( -s\ln \hat{H}%
\right) ,s\in C.\eqno(A.1)$$
In the case where we have a known spectrum, $\left\{ \lambda _n\right\} $ (
for simplicity, we assume in what follows that the spectrum is discrete),
the operator $e^{-s\ln \hat{H}}$ may be diagonalized, becoming an
infinite ``matrix'', with entries $e^{-s\ln \lambda _n}$, and therefore the
trace in Eq.(A.1) becomes: 
$$\zeta _{\hat{H}}\left( s\right) \equiv \sum_n\lambda _n^{-s}.\eqno(A.2)$$
Of course, if one has a zero eigenvalue, it is dealt with separately. If $%
\lambda _n=n$, where $n$ is a positive integer, and the summation runs over
the positive integers, one has just the well known Riemann zeta function
~\cite{Gra}, and hence its name. Even if the spectrum is not known, one can still
obtain information about the $\zeta $-function by the so-called heat
equation of the operator $\hat{H}$:%
$$\frac \partial {\partial t}F(x,y,t)+\hat{H}F(x,y,t)=0,\eqno(A.3)$$
where $x$ and $y$ are spacetime points, and $t$ is a parameter not
necessarily equal to the coordinate time. We assume that $\hat{H}$ acts
with respect to $x$ only, and that we have the initial condition 
$$F(x,y,0)\equiv \delta (x-y).\eqno(A.4)$$
Of course, Eq.(A.3) reduces to the ordinary heat equation if we have $%
\hat{H}=-\Delta $, the ordinary Laplacian. The function $F(x,y,t)$
(actually a distribution) obeying Eq.(A.3) with condition as Eq.(A.4) is
called the {\it heat kernel} of the operator $\hat{H}$.

Now let $\left\{ \varphi _n\left( x\right) \right\} $ be a complete set of
eigenfunctions of $\hat{H}$, orthonormalized by the standard inner
product: 
$$\int \varphi _n\left( x\right) \varphi _m\left( x\right) \sqrt{g}dx=\delta
_{mn},\eqno(A.5)$$ 
where $g$ is the determinant of the metric tensor on the spacetime, or
region of spacetime under consideration. Of course, they have associated
eingenvalues: 
$$\hat{H}\varphi _n=\lambda _n\varphi _n.\eqno(A.6)$$ 
In terms of these eigenvalues and eigenfunctions the heat kernel may be
written as: 
$$F(x,y,t)=\sum_n\exp (-\lambda _nt)\varphi _n(x)\varphi _n(y).\eqno(A.7)$$ 

The trace of the heat kernel is defined as: 
$$Tr\exp (-t\hat{H)}\equiv \int F(x,x,t)\sqrt{g}dx=\sum_n\exp (-\lambda
_nt).\eqno(A.8)$$ 

Note that $F(x,y,t)$ is the ``matrix ''element of the operator $\exp (-t%
\hat{H})$. The link with the $\zeta $-function is established via a
Mellin transform~\cite{Haw}: 
$$\zeta _{\hat{H}}(s)=\frac 1{\Gamma \left( s\right) }\int_0^\infty
t^{s-1}Tr\exp (-t\hat{H})dt.\eqno(A.9)$$ 

One may also define the {\it local} $\zeta $-function : 
$$\zeta _{\hat{H}}(x,s)=\frac 1{\Gamma \left( s\right) }\int_0^\infty
t^{s-1}F(x,x,t)dt,\eqno(A.10)$$ 
in terms of which one may adapt Eq.(5) to obtain the free energy{\it \
density:} 
$${\cal F}=\frac 12\zeta _{\hat{H}}(x,-1/2)-\frac 1{\sqrt{4\pi }%
}\sum_{n=1}^\infty \int_0^\infty t^{-3/2}e^{-(n\beta )^2/4t}F(x,x,t)dt.\eqno(A.11)$$ 

The convenience of taking local quantities is avoiding an infinite global $%
\zeta $-function due to an infinite spacetime.
\\
\\
{\large{\bf Appendix B: the heat kernel for the screw dislocation}}
\\
\\
In Section III, we displayed the heat kernel for the operator $L_3=-\Delta
_{LB}+m^2$ for the metric describing the screw dislocation, Eq.(6). We also
showed in Appendix A that the problem of obtaining the heat kernel actually
boils down to knowing the eingenvalues and (orthonormalized) eigenfunctions
of the respective operator, and using Eq.(A.7). The eigenvalue equation can
be written for the screw dislocation metric as: 
$$\left[ -\frac 1r\frac \partial {\partial r}\left( r\frac \partial {\partial
r}\right) -\frac 1{r^2}\left( \frac \partial {\partial \phi }-\kappa \frac
\partial {\partial z}\right) ^2-\frac{\partial ^2}{\partial z^2}+m^2\right]
\varphi \left( \vec{r}\right) =\lambda \varphi \left( \vec{r}%
\right) .\eqno(B.1)$$
A complete set of orthonormal solutions is easily seen to be: 
$$\varphi _{n,\nu ,k}\left( r,\phi ,z\right) =\frac 1{2\pi }J_{\mid n+\kappa
\nu \mid }(kr)e^{-in\phi }e^{i\nu z},\eqno(B.2)$$
and its complex conjugate. $\lambda =k^2+\nu ^2+m^2$, where $k$ is a
positive real number, $\nu $ is a real number, and $n$ is an integer. $%
J_{\mid n+\kappa \nu \mid }(kr)$ is the regular Bessel function. Application
of Eq.(A.7) gives (the measure of integration for $k$ is $kdk$):
$$\begin{array}{rcl} 
K(\vec{r},\vec{r}^{\prime },t)&=&\frac{e^{-m^2t}}{\left( 2\pi
\right) ^2}\sum_{n=-\infty }^\infty \int_{-\infty }^\infty d\nu \ e^{-\nu
^2t}\int_0^\infty dk\ ke^{-k^2t}e^{-in\left( \phi -\phi ^{\prime }\right)
}e^{i\nu \left( z-z^{\prime }\right) }\times\\[0.3cm]
&&J_{\mid n+\kappa \nu \mid }(kr)J_{\mid
n+\kappa \nu \mid }(kr^{\prime }).
\end{array}\eqno(B.3)$$
Using the relation ~\cite{Gra}: 
$$\int_0^\infty e^{-a^2x^2}J_p\left( \alpha x\right) J_p\left( \beta x\right)
xdx=\frac{e^{-\frac{\left( \alpha ^2+\beta 2\right) }{4a^2}}}{2a^2}I_p\left( 
\frac{\alpha \beta }{2a^2}\right) ,\eqno(B.4)$$
we arrive at the form: 
$$K\left( \vec{r},\vec{r}^{\prime },t\right) =\frac{e^{-m^2t}e^{-%
\frac{\left( r^2+r^{\prime 2}\right) }{4t}}}{2\left( 2\pi \right) ^2t}%
\sum_{n=-\infty }^\infty e^{-in\left( \phi -\phi ^{\prime }\right)
}\int_{-\infty }^\infty d\nu \ e^{-\nu ^2t}e^{i\nu \left( z-z^{\prime
}\right) }I_{\mid n+\kappa \nu \mid }\left( \frac{rr^{\prime }}{2t}\right) .\eqno(B.5)$$

In order to show that it is consistent, we put $b=0$ to show it reduces to
the Euclidean case. Using the identity~\cite{Gra}: 
$$e^{\frac x2\left( t+\frac 1t\right) }=\sum_{n=-\infty }^\infty I_n\left(
x\right) t^n,\eqno(B.6)$$
we obtain: 
$$K_{b=0}(\vec{r},\vec{r}^{\,\prime },t)=\frac{e^{-m^2t}e^{-\frac{%
\left( \vec{r}-\vec{r}^{\prime }\right) ^2}{4t}}}{\left( 4\pi
t\right) ^{3/2}},\eqno(B.7)$$
which in fact is the Euclidean heat kernel~\cite{Las}.
\\
\\
{\Large{\bf Acknowledgment}}

\noindent
This work was partially supported by CNPq.

\end{document}